# Thermo-physical description of elastic solids

József Garai

*Department of Earth Sciences, Florida International University, Miami, FL 33199, USA*

Experiments show that all the derivatives of the thermo-physical variables are nearly constant. The constant value of the derivatives indicates linear relationship between the variables. Neither the volume coefficient of thermal expansion nor the bulk modulus are constant and show pressure and temperature dependence. Besides the derivatives the only other quantity in the definition of the volume coefficient of thermal expansion and the bulk modulus is the volume. It is suggested that the pressure and temperature dependence of these parameters is resulted from the improper identification of the volume. In solid phase the actual volume is not an independent variable but rather comprises from three more fundamental volume components, initial, thermal, and elastic volumes. The pressure correlates to the elastic volume and the temperature to the thermal volume. New definitions for the volume coefficient of thermal expansion and the bulk modulus are proposed by identifying the fundamental volume parts specifically. The complete separation of the thermal and elastic volumes is consistent with theory since the vibrational and static energies are completely independent of each other. Based on the independence of the thermal and elastic deformations it is suggested that thermo elastic coupling should not exist in solid phase in the elastic domain with the exception of constant volume. It is also suggested that the thermo-elastic coupling is irreversible and exist only in the temperature-pressure direction.

# I. INTRODUCTION

In solid phase the relationships among the thermo-physical variables, pressure [p], temperature [T], and volume [V] are defined by the volume coefficient of expansion [$\alpha_V$] and the bulk modulus [$B_T$]. The volume coefficient of expansion describes the relationship between temperature and volume

$$\alpha_{Vp} \equiv \frac{1}{V}\left(\frac{\partial V}{\partial T}\right)_p \tag{1}$$

while the correlation between the pressure and volume is given by the isothermal bulk modulus:

$$B_T \equiv -V\left(\frac{\partial p}{\partial V}\right)_T \tag{2}$$

For the validity of equations (2) it is assumed that the solid is homogeneous, isotropic, non-viscous and has linear elasticity. It is also assumed that the stresses are isotropic; therefore, the principal stresses can be identified as the pressure[1] $p = \sigma_1 = \sigma_2 = \sigma_3$.

Both the volume coefficient of expansion and the isothermal bulk modulus are pressure and temperature dependent; therefore, the universal description of solids requires knowing the derivatives of these parameters.

$$\left(\frac{\partial \alpha_V}{\partial T}\right)_p ; \quad \left(\frac{\partial \alpha_V}{\partial p}\right)_T ; \quad \left(\frac{\partial B_T}{\partial T}\right)_p ; \quad \left(\frac{\partial B_T}{\partial p}\right)_T \tag{3}$$

# II. FUNDAMENTAL COMPONENTS OF THE VOLUME

Contrarily to gasses Avogadro's principle does not apply to solids. Matter in solid phase occupies an initial volume [$V_o$] at zero pressure and temperature.

$$V_o = nV_o^m, \tag{4}$$

In Eq. (4) n is the number of moles and $V_o^m$ is the molar volume of the substance at zero pressure and temperature. The pressure modifies this initial volume by inducing elastic deformation while the temperature by causing thermal deformation. Using equations (1) and (2) the actual volume at given pressure and temperature can be calculated[2] by allowing one of the variables to change while the other one held constant

$$[V_T]_{p=0} = V_o e^{\int_{T=0}^{T} \alpha_{p=0} dT} \quad \text{or} \quad [V_p]_{T=0} = V_o e^{-\int_{p=0}^{p} \frac{1}{B_{T=0}} dp} \tag{5}$$

and then

$$[V_T]_p = [V_p]_{T=0} e^{\int_{T=0}^{T} \alpha_{V_p} dT} \quad \text{or} \quad [V_p]_T = [V_T]_{p=0} e^{-\int_{p=0}^{p} \frac{1}{B_T} dp} \tag{6}$$

These two steps might be combined into one and the volume at a given p, and T can be calculated:

$$V_{p,T} = V_o e^{\int_{T=0}^{T} \alpha_{V_{p=0}} dT - \int_{p=0}^{p} \frac{1}{B_T} dp} = V_o e^{\int_{T=0}^{T} \alpha_{V_p} dT - \int_{p=0}^{p} \frac{1}{B_{T=0}} dp} \tag{7}$$

The total volume change related to the temperature will be called thermal volume $[V^{th}]$ while the total volume change resulted from elastic deformation will be called elastic volume $[V^{el}]$. The thermal volume at zero pressure is

$$[V_T^{th}]_{p=0} = V_o \left( e^{\int_{T=0}^{T} \alpha_V dT} - 1 \right), \tag{8}$$

while the elastic volume at zero temperature is

$$[V_p^{el}]_{T=0} = V_o \left( e^{-\int_{p=0}^{p} \frac{1}{B_T} dp} - 1 \right). \tag{9}$$

The thermal volume at pressure p is

$$[V_T^{th}]_p = [V_T^{th}]_{p=0} e^{-\int_{p=0}^{p}\frac{1}{B_T}dp} = V_o e^{-\int_{p=0}^{p}\frac{1}{B_T}dp}\left(e^{\int_{T=0}^{T}\alpha_V dT} - 1\right), \qquad (10)$$

and the elastic volume at temperature T is

$$[V_p^{el}]_T = [V_p^{el}]_{T=0} e^{\int_{T=0}^{T}\alpha_V dT} = V_o e^{\int_{T=0}^{T}\alpha_V dT}\left(e^{-\int_{p=0}^{p}\frac{1}{B_T}dp} - 1\right). \qquad (11)$$

The actual volume is the sum of the volume components:

$$[V_T]_p = V_o + [V_p^{el}]_{T=0} + [V_T^{th}]_p \quad \text{or} \quad [V_p]_T = V_o + [V_p^{el}]_T + [V_T^{th}]_{p=0} \qquad (12)$$

Since

$$[V_T]_p = [V_p]_T \qquad (13)$$

from Eq (12) follows that

$$[V_T^{th}]_p - [V_T^{th}]_{p=0} = [V_p^{el}]_T - [V_p^{el}]_{T=0}. \qquad (14)$$

The compressed part of the thermal volume is the same as the expanded part of the elastic volume. Since the volume difference in Eq. (14) both temperature and pressure dependent I will call this volume difference to thermo-elastic volume $[\Delta V_p^{th-el}]_T$

$$[V_p^{th-el}]_T = [V_T^{th}]_p - [V_T^{th}]_{p=0} = [V_p^{el}]_T - [V_p^{el}]_{T=0}. \qquad (15)$$

The thermoelastic volume can be calculated as:

$$[V_p^{th-el}]_T = V_o\left(e^{\int_{T=0}^{T}\alpha_V dT} - 1\right)\left(e^{-\int_{p=0}^{p}\frac{1}{B_T}dp} - 1\right). \qquad (16)$$

It can be concluded that the actual volume comprises from four distinct volume parts, initial volume, thermal volume at zero pressure $[V_T^{th}]_{p=0}$, elastic volume at zero temperature $[V_p^{el}]_{T=0}$, and thermo-elastic volume $[V_p^{th-el}]_T$ (Fig. 1).

$$V = V_o + [V_T^{th}]_{p=0} + [V_p^{el}]_{T=0} + [V_p^{th-el}]_T. \tag{17}$$

These fundamental volume components are related to the thermo-physical variables as:

$$V_o = f(n), \tag{18}$$

$$[V_T^{th}]_{p=0} = f(n,T), \tag{19}$$

$$[V_p^{el}]_{T=0} = f(n,p), \tag{20}$$

and

$$[V_p^{th-el}]_T = f(n,T,p). \tag{21}$$

It can be concluded that the volume is not an independent variable but rather the sum of four more fundamental volume parts. Describing the thermo-physical relationships between the volume and the pressure and between the volume and the temperature require the proper identification of the appropriate volume part. The improper definition of the volume in the thermo-physical relationships most likely skipped the attention of researchers for centuries since the integral of the volume

$$\int \frac{1}{x} dx = \ln(x) + c. \tag{22}$$

gives the same result as the integral of the specifically defined volume

$$\int \frac{1}{a+x} dx = \ln(a+x) + c. \qquad (23)$$

However, the mathematical equivalency of the integrations in Eq. (22) and (23) can not guarantee that equivalency between these equations are exist for other mathematical operations. Thus the fundamental volume must be correctly identified in the thermo-physical relationships.

### III. LINEAR RELATIONSHIPS AMONG THE THERMO-PHYSICAL VARIABLES

Experiments indicate that the product of the volume coefficient of thermal expansion and the bulk modulus remains almost constant at temperatures higher than the Debye temperature[3-6].

$$[\alpha_V B_T = \text{const}]_{T>\theta_D} \qquad (24)$$

The relatively minor deviation[7] of $\alpha_{V_p} K_T$ from constant value can be described as:

$$\alpha_{V_p} K_T = m\left(1 - e^{-\frac{T}{n}}\right) \qquad (25)$$

where m and n are constant for a given solid.

Assuming that the product of the volume coefficient of thermal expansion and the bulk modulus is constant it can be written that

$$\alpha_{V_{p=0}} K_{T=0} = \alpha_{V_{p=0}} K_{T=T} = \alpha_{V_{p=p}} K_{T=0} = \text{const}. \qquad (26)$$

Based on Eq. (26) it can be shown that

$$\left[\frac{\partial V^{th}}{\partial T} = \text{const}\right]_{T>\theta_D} \qquad \text{and that} \qquad \left[\frac{\partial p}{\partial V^{el}} = \text{const}\right]_{T>\theta_D}. \qquad (27)$$

At constant volume the differentials of the elastic and thermal volumes are equal

$$\partial V^{el} = \partial V^{th} \qquad (28)$$

resulting that differential of the pressure and the temperature also remains constant

$$\frac{\partial p}{\partial T} = \text{const}.\qquad(29)$$

The constant value of the product of the volume coefficient of thermal expansion and the bulk modulus indicates that all the derivatives of the thermo-physical variables remain constant. Regardless of the constant value of the derivatives both the volume coefficient of thermal expansion and the bulk modulus are pressure and temperature dependent. Besides the derivatives of the variables the only additional quantity in these expressions is the volume. I interpret this fact as an indication the volume in the definition of the volume coefficient of thermal expansion and the bulk modulus has not been identified correctly. The incorrect definition of the volume introduces an artificial term, which alters the originally linear relationship of the variables to a nonlinear one.

Both the pressure and the temperature derivatives of the volume coefficient of thermal expansion and the bulk modulus are nonzero indicating that the artificial part in the volume should be pressure and temperature dependent. The only volume component, which both pressure and temperature dependent is the thermo-elastic volume [Eq. (21)] indicating that this volume part alters the originally linear relationships of the thermo-physical variables to non linear.

The assumption, the thermo-elastic volume alters the linear relationship of the variables, is supported by the observed constant derivative of the pressure and the temperature in Eq. (29). At constant volume the thermal deformation completely converted to elastic deformation. As a result neither the thermal nor the elastic volume contains contribution from the thermo-elastic volume (Fig. 1). The elimination of the thermoelastic volume results that both the volume coefficient of thermal expansion and the bulk modulus remains constant.

## IV. PROPOSED NEW DEFINITIONNS FOR THE VOLUME COEFFICIENT OF THERMAL EXPANSION AND BULK MODULUS

Deducting the thermal-elastic volume from the total volume gives

$$V - [V_p^{th-el}]_T = V_o \left( e^{\int_{T=0}^{T} \alpha_V dT} + e^{-\int_{p=0}^{p} \frac{1}{B_T} dp} - 1 \right). \quad (30)$$

The elimination of the thermoelastic volume requires the transformation of the $V - [V_p^{th-el}]_T$ volume to the actual volume V:

$$V - [V_p^{th-el}]_T = V_o \left( e^{\int_{T=0}^{T} \alpha_V dT} + e^{-\int_{p=0}^{p} \frac{1}{B_T} dp} - 1 \right) \Rightarrow V. \quad (31)$$

Assuming that the derivatives of the variables remain constant, the transformation can be achieved by redefining the volume coefficient of thermal expansion and the bulk modulus

$$\alpha_{V_p} \Rightarrow \alpha_V^o \qquad B_T \Rightarrow B^o. \quad (32)$$

Superscript o is used for the new parameters which are defined as:

$$\alpha_V^o \equiv \frac{1}{V_{p=0}} \frac{\partial V^{th}}{\partial T} \quad (33)$$

and

$$B^o \equiv -V_{T=0} \frac{\partial p}{\partial V^{el}}. \quad (34)$$

Using the new definition of the volume coefficient of thermal expansion the thermal volume is

$$V_o^{th} = V_o \left( e^{\int_{T=0}^{T} \alpha_V^o dT} - 1 \right) \quad (35)$$

while the elastic volume can be calculated as:

$$V_o^{el} = V_o \left( e^{-\int_{p=0}^{p} \frac{1}{B^o} dp} - 1 \right). \tag{36}$$

Subscript o is used to show that these fundamental volume parts were determined by using $B_o$ and $\alpha_o$. The actual volume is the sum of the initial, thermal and elastic volumes

$$V = V_o + V_o^{el} + V_o^{th}. \tag{37}$$

Substituting the fundamental volume components gives the actual volume

$$V = V_o \left( e^{-\int_{p=0}^{p} \frac{1}{B^o} dp} + e^{\int_{T=0}^{T} \alpha_V^o dT} - 1 \right). \tag{38}$$

Eq. (38) is identical with the required expression given in Eq. (31) except the volume coefficient of expansion and the bulk modulus have been determined in accordance to the new definitions. Thus the transformation of the volume is completed through the introduction of the new definitions [Eq. (33) and (34)]. The universal EOS of elastic solids [Eq. (38)] describes the relationships among the thermo-physical variables pressure, volume, and temperature.

## V. THERMO-ELASTIC COUPLING

In equations (33) and (34) the thermal and elastic related deformations are entirely separated. This complete separation is supported by theory. The internal energy near the classical limit $T/\theta = 0.8$ is given as

$$U = E_{zp} + E_0 + E_{th}. \tag{39}$$

where $E_{zp}$ is the zero-point vibrational energy, $E_0$ is the static energy of the lattice at absolute zero, and $E_{th}$ is the thermal energy arising from the motion of the atoms about the lattice point positions. Neglecting the zero-point energy the pressure is given as:

$$p = \frac{\partial E_0}{\partial V^{el}} + \frac{\partial E_{th}}{\partial V^{th}}. \quad (40)$$

Increasing the temperature at zero pressure results that the static energy is zero, while the differentials of the thermal energy and the thermal volume are non zero

$$\frac{\partial E_0}{\partial V^{el}} = 0 \quad ; \quad \frac{\partial E_{th}}{\partial V^{th}} \neq 0 \quad \text{and} \quad p = 0. \quad (41)$$

The non-zero value of the derivative of the thermal volume at zero pressure contradict with Eq. (40) and prove that the thermal energy and the thermal volume do not correlate to the pressure. By replacing the differentials of the thermal volume with the elastic volume Eq. (40) must be rewritten as:

$$p = \frac{\partial E_0}{\partial V^{el}} + \frac{\partial E_{th}}{\partial V^{el}}. \quad (42)$$

The independence of the vibrational and static energies is shown in a one dimensional toy model (Fig. 2). Eventhough, the vibrational and the static energies are independent, the pressure and the temperature correlates to each other at constant volume because at this condition the thermal volume is forced to convert to elastic volume. The relationship is irreversible, which can be demonstrated by calculating the mechanical energy stored by the system. Changing the pressure modifies the elastic volume and produces work. The work between the initial and final state can be approximated as the area of a trapezoid (Fig. 3)

$$(\Delta w_{i-f})_T \approx \int_i^f \left( p_i + \frac{\partial p}{2} \right) \partial V^{el}. \quad (43)$$

Subscript i and f are used for the initial and final conditions respectively. In order to change the thermal volume first the elastic volume should be restored into its initial state. Restoring the initial elastic volume requires the same mechanical energy with opposite sign which already supplied to the system

$$(\Delta w_{i-f})_T = -(\Delta w_{f-i})_T. \tag{44}$$

Since all the energy is used up to restore the initial volume there is no energy left which could be transformed into thermal energy. The energy equivalence between the initial and the volume restoration energies proves that the pressure change will not change the thermal volume; consequently the temperature will remain the same. This conclusion is consistent with the complete recovery nature of the elastic energy.

The irreversible relationship between the temperature and the pressure is true only for non-viscous elastic solids. Viscous or plastic deformations are partially or fully reversible. The relationship between the thermo-physical quantities is shown in FIG. 4.

The separation of the static and vibrational energies along with their related volume changes results that the mechanical work for elastic solids (s) must be written as

$$\delta w(s) = \delta w(s)^{el} \tag{45}$$

and for isothermal work

$$[\delta w(s)]_T = -p(dV^{el}_{p/V})_T \tag{46}$$

The thermodynamic condition, no mechanical work is done on/by the system, is given by the condition $\partial V = 0$. This condition is correct only for gas phase. For elastic solids the conditions must be given as $\partial V^{el} = 0$. Like the traditional definition of the heat capacity is correct only for gasses (g)

$$[c(g)]_V = \left(\frac{\partial U}{\partial T}\right)_V \tag{47}$$

for solids the heat capacity is defined as:

$$[c(s)]_{V^{el}} = \left(\frac{\partial U}{\partial T}\right)_{V^{el}}. \tag{48}$$

## VI. CONCLUSIONS

The volume is not a fundamental quantity but rather comprises from three different parts, initial, thermal, and elastic volumes. The thermal and elastic volumes are the result of temperature and pressure respectively. It is suggested that the appropriate fundamental volume component must be identified specifically in the thermo-physical relationships. According to this criterion new definitions for the volume coefficient of expansion and the bulk modulus are proposed. The new definitions of these parameters results a new p-V-T equation of state for elastic solids. The specific evaluation of the proposed parameters and the new EoS to experiments is not the subject of this study and will be discussed in a separate paper.

The proposed new definitions for the volume coefficient of expansion and the bulk modulus are consistent not only with experiments but also with theory. The independent physical processes, vibrational and static, result that no thermo-elastic coupling can exist in solid phase. The only exception is the constant volume condition when the thermal volume is forced to convert to elastic volume. The relationship between the temperature and pressure at this condition is irreversible since the pressure can not generate heat. The lack of thermo-elastic coupling means that adiabatic conditions should not exist in solid phase in the elastic domain.

## ACKNOWLEDGEMENT

I would like to thank Alexandre Laugier for his encouragement and helpful comments on the manuscript.

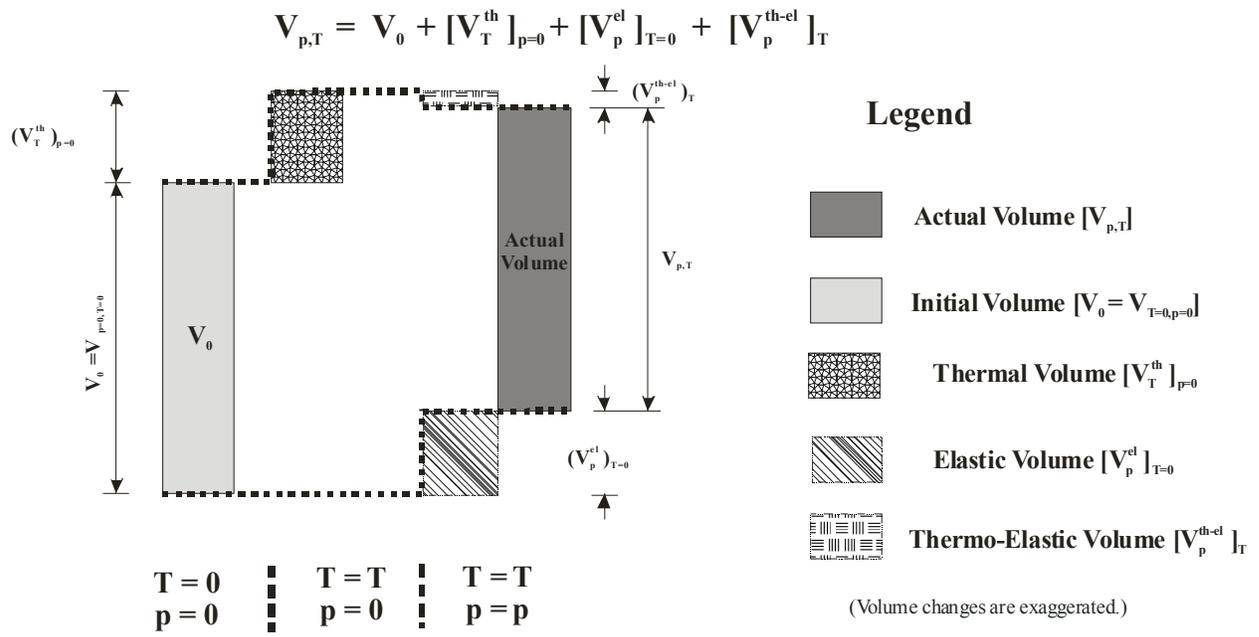

FIG. 1 The fundamental volume components of the actual volume, in accordance to the conventional thermo-physical description of solids.

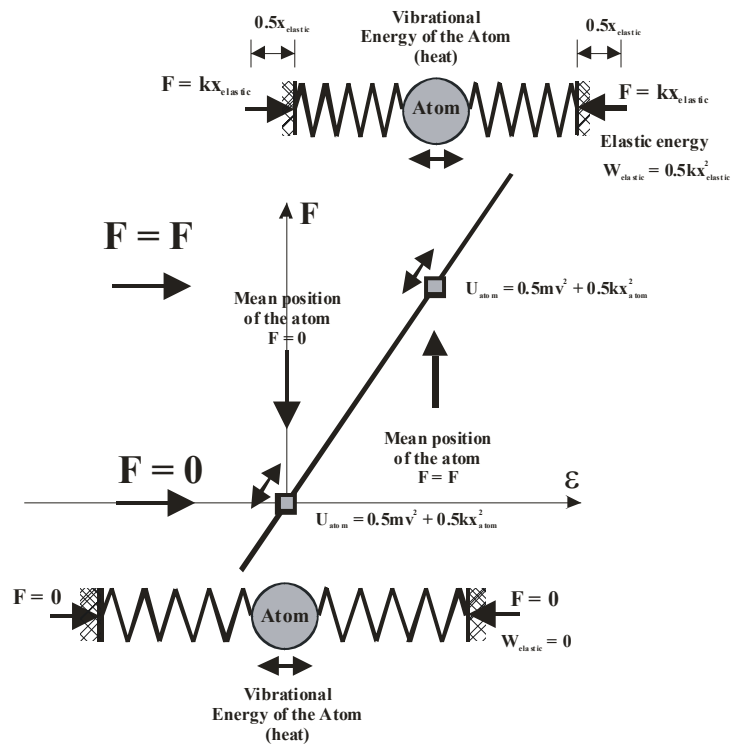

FIG. 2. One dimensional toy model demonstrating the independence of the vibrational (thermal) and static (elastic) energies. It can be seen that the vibration of the atom is not effected by the stress exist in the spring if the elasticity is linear.

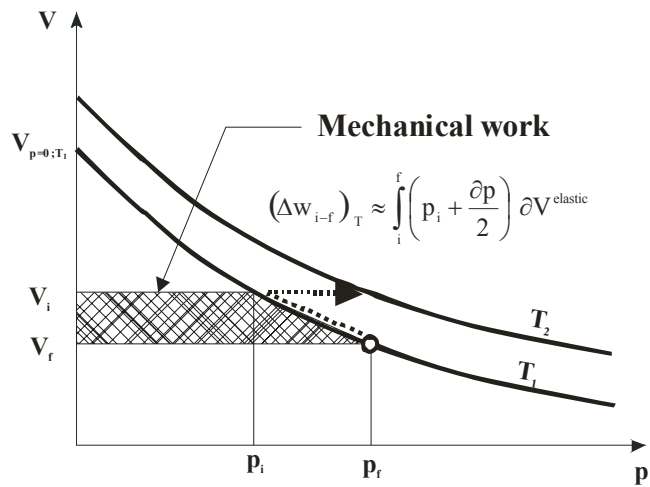

FIG. 3. The work done on the system between the initial and final condition. The restoration of the volume uses up all the energy leaving no energy left for thermal volume change and for modifying the temperature.

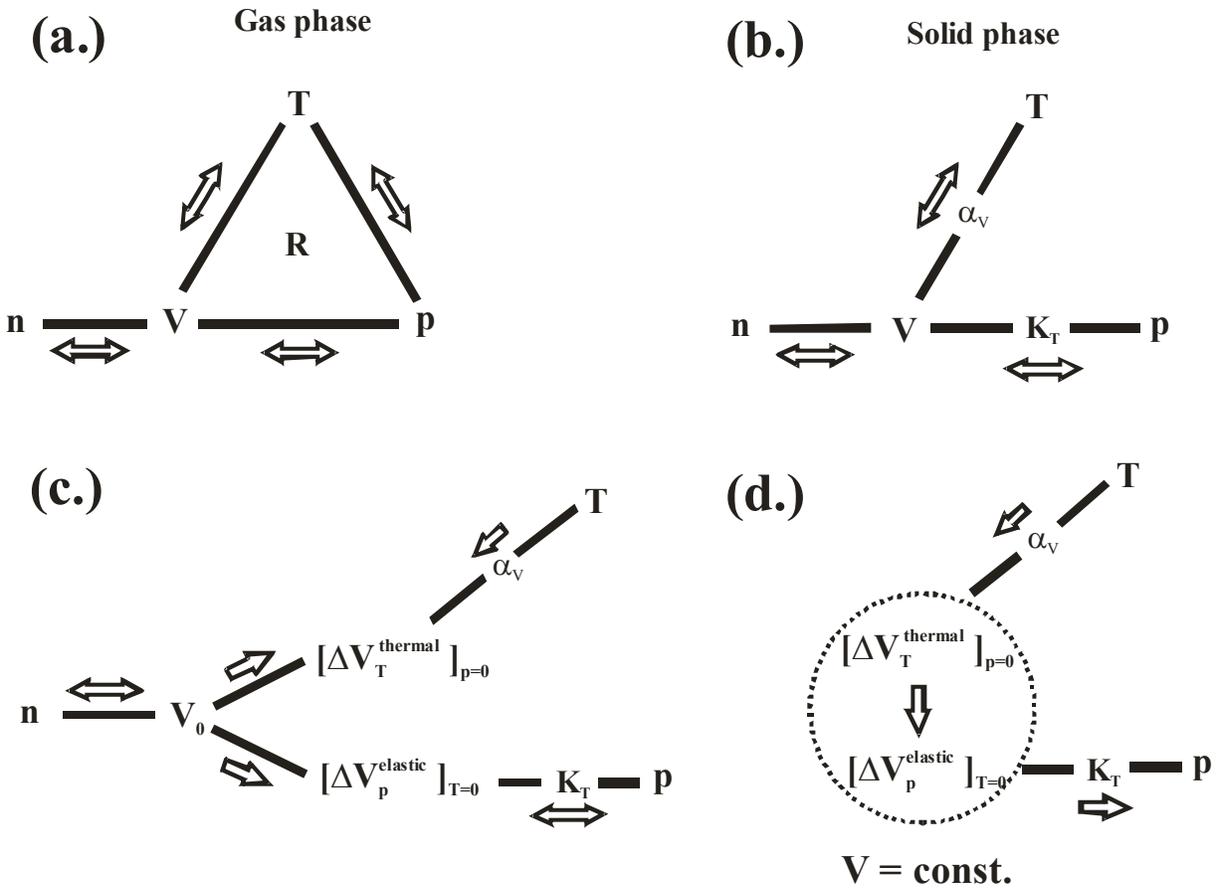

FIG. 4. Thermo-physical relationships (a) gas phase, (b) solid phase conventional description (c) proposed new description (d) constant volume. (The arrow ↔ represent a reversible while → represents an irreversible relationship or process.)